\documentclass{an}
\usepackage{graphicx}
\usepackage{times}
\usepackage{fancyhdr}
\newcommand{\drv}[2]{\frac{\partial #1}{\partial #2}   }

\newcommand{\ddrv}[2]{\frac{\partial^{2}  #1}{{\partial #2}^2}   }

\begin{document}

\include{inc_header}

\title{Resonance conditions}

\author{P. Rebusco}
\institute{Max-Planck-Institut f\"ur Astrophysik, 85741 Garching, Germany}

\date{Received 8 September 2005; accepted 22 September 2005; 
published online 20 October  2005}

\abstract{
Non-linear parametric resonances occur frequently in nature.
Here we summarize
how they can be studied by means of perturbative methods. We show in particular
 how resonances can affect the motion of a test particle orbiting in the
vicinity of a compact object.
These mathematical toy-models find application in explaining the structure
of the observed kHz Quasi-Periodic Oscillations: we show which aspects of
the reality naturally enter in the theory, and which one  still remain a puzzle. 
}

\correspondence{pao@mpa-garching.mpg.de}

\maketitle

\section{Introduction}
A particular analytic model for the Klu\'zniak-Abramowicz resonance idea for Quasiperiodic 
Oscillations (QPOs) was
developed by Rebusco (2003) and Hor\'ak (2004).
It describes the QPO phenomenon in terms of two coupled non-linear forced oscillators,
\begin{eqnarray}
      \label{eqns}
   \ddot\delta r+\omega_r^2 \delta r &=& F(\delta r,\delta z, \dot\delta r, \dot \delta z)+A \cos{(\omega_0 t)}+N_r(t),\\
       \ddot\delta z+\omega_{z}^2 \delta z &=&  G(\delta r,\delta z, \dot\delta r, \dot \delta z)+B \cos{(\omega_0 t)}+N_z(t),\nonumber
\end{eqnarray}
where $F$ and $G$ are polynomials of second  or higher degree (obtained in terms of expansion of
 the deviations from a Keplerian flow) and $\omega_r$ and
 $\omega_z$ are the epicyclic frequencies (see next section). 
 The $\cos(\omega_0 t)$ terms 
  represent an external forcing: they are mostly important in the case of Neutron Stars (NS), 
  where $\omega_0$
is the NS spin frequency. $N_r$ and $N_z$ describe the stochastic noise due to
the magneto-rotational instability (MRI).
Till now, solutions of (\ref{eqns})  have been studied in detail in the case of unforced
    oscillations without noise.  In what follows we describe shortly these solutions, and comment on astrophysical
        consequences.

\section{Nearly Keplerian motion}
\label{keplerian}
\subsection{Perturbed test particle}
We analyze the case of a single test particle moving in a strong gravitational field.   
Consider a free test particle orbiting in the vicinity of a nonrotating
compact object: it moves along the geodesics given by the Schwarzschild metric
\begin{eqnarray}
\ddot \theta(\tau) & + & \frac{E^2}{2 g_{\theta\theta}}
 \drv{U_{\rm eff}(r,\theta)}{\theta}+2 \Gamma_{r \theta}^{\theta}\dot r(\tau) \dot \theta(\tau)=0\:, \\
\ddot r(\tau)& + &\frac{E^2}{2 g_{rr}}
 \drv{U_{\rm eff}(r,\theta)}{r}+ \Gamma_{\theta \theta}^{r} {\dot \theta(\tau)}^2+\Gamma_{rr}^{r} {\dot r(\tau)}^2=0\:, \nonumber \\
\ddot\phi(\tau) &= &0\:, \nonumber 
\label{eqGR}
\end{eqnarray}
where $U_{\rm eff} = g^{tt} +l g^{t \phi}+l^2 g^{\phi \phi}$ is the effective
potential  and $E=-u_t$ is the energy, $(\theta,r,\phi)$ are the spherical components.
Let us slightly perturb the orbit of such a particle: what happens?
In first approximation 
 the particle  oscillates harmonically in each direction of the perturbation; to higher orders
however the different directions are coupled and nonlinear effects show up.
An example of such a perturbation was studied numerically by Abramowicz et al.\ (2003) 
and analytically by Rebusco (2004): these are specific examples
(Taylor expansion to third order plus
introduction of an arbitrary constant $\alpha$), but they display a behaviour which is
     common to any nearly-Keplerian motion. 
  In general a perturbation in the radial ($\delta r$) and in the vertical ($\delta z$) direction can be written in the  form (\ref{eqns}).
 Up to now a study of these equations (without the turbulent noise terms $N$)
have been done by using perturbative methods; among them, we found that the method of 
multiple scales (read Sect.~3) is particularly useful.

\subsection{Particle motion and Eigenfrequencies} 
The radial epicyclic frequency of planar motion and the vertical epicyclic
 frequency of nearly off-plane motion are respectively defined as
\begin{eqnarray}
\omega_r^2 &=& \left (\frac{1}{g_{rr}}\ddrv{U_{\rm eff}}{r}\right )_{\!\ell,r_0,\pi/2}\ ,\\
\omega_{z}^2 &=& 
\left (\frac{1}{g_{\theta\theta}}\ddrv{U_{\rm eff}}{\theta}\right )_{\!\ell,r_0,\pi/2}\ .
\nonumber
\end{eqnarray}
They depend only on the metric of the system, hence on strong gravity itself.
Moreover they scale from source to source with $M^{-1}$ , giving a unique possibility to 
fix the mass $M$ of the central object, once their relation with the observed QPOs is clarified. 
 For a spherically symmetric gravitating fluid body (a better model of the accretion
  disk) these are the frequencies at which the center of mass 
  (initially on a circular geodesic) oscillates (Abramowicz et al.\ 2005): hence the
   toy-model of a single particle is meaningful also for  more complex systems, in
    which the coupling arises naturally.

In Newtonian gravity there is degeneracy between these eigenfrequencies and
 the Keplerian frequency: that is why bounded orbits are closed.
  In general relativity this degeneracy is broken and as a consequence two (for nonrotating object)
   or three (for Kerr's BH or NS) characteristic frequencies are present, opening the possibility 
   of internal resonances (see Fig.~2 in Abramowicz et al.\ 2005).  
External resonances are likely to occur
     too: this is the well known case of the rings of Saturn, in which the structure of the gaps  results from resonance with outer satellites. This could be the case of a
      perturbation at the stellar spin frequency in NSs: a nonlinear forced resonance would  explain why in many NSs the frequency difference of the two
peaks remains close to $\frac{1}{2}$ the spin, while the position of the two peaks
varies in time (van der Klis $2000$). 

 The resonant model (see the review by Klu\'zniak 2005) is based on such observations and on the
  analogy of the perturbed equations with the Mathieu equation, which describes a swing with
   oscillating point of suspension. In the next section we will see why the $3:2$ ratio
    is the most likely  to take place and why the observed frequencies are near 
     the epicyclic ones, but not exactly  the same.

\section{Weakly non-linear oscillators}
Weakly anharmonic oscillators can be approximated by harmonic
 oscillators whose frequencies are close to the eigenfrequencies. This fact
  was already well known by XIX century astronomers; Henry Poincar\'e developed
   a method (e.g.\ Poincar\'e $1882$) to study these systems, based on the assumption 
   that the approximate solution is in the form $x_i=a_i \cos{\omega_i^* t+\phi_i}+\Sigma_j \epsilon^j x_i^{(j)}$, where
     $\omega_i^*=\omega_i+\epsilon \omega_i^{(1)}+\epsilon^2 \omega_i^{(2)}+ \epsilon^3 \omega_i^{(3)}+...$ and $\omega_i$ are the eigenfrequencies
     ($i=1,..,n$, with $n$ degrees of freedom).
     The frequency corrections $\omega_i^{(j)}$ are found to depend on the amplitudes of
      the perturbation and on the constants of the system. $\epsilon$ is a constant $\ll1$,
       which measures the deviation from the linearity: its value has to be suitably
        linked to some physical quantities, but here it is sufficient to notice
 that it is small.

 Another method was developed
 more recently (e.g.\ Mook \& Nayfeh $1976$), the so-called method of multiple scales,
which permits to get more information about such systems (see Hor\'ak 2004 for a
review).  This method is another variation of the straightforward expansion, whose 
fundamental idea is to consider the expansion which
represents the approximate solution to be a function of multiple
independent time scales, instead of a single time.
The new ``time-like'' independent variables are defined:
$T_k = \epsilon^k t\:\: \:\mbox{for}\:\:\: k=0,1,2...\,$.
Expressing the solution as a function of more variables, treated as
 independent, is a trick to remove the terms which would
make the solution unphysically to diverge.
 By writing $T_k$ one makes a formal assumption  of physical phenomena which
  vary at slower time scales: indeed the different nonlinear effects accumulate in 
  slower time scales and with these expansions we are able to study them separately.

The term parametric resonance is used to describe not closed oscillating
 systems in which the external action vary the parameters of the system itself
  (e.g.\ Mathieu equation).  
\begin{figure}
\resizebox{\hsize}{!}
{\includegraphics[scale=0.5]{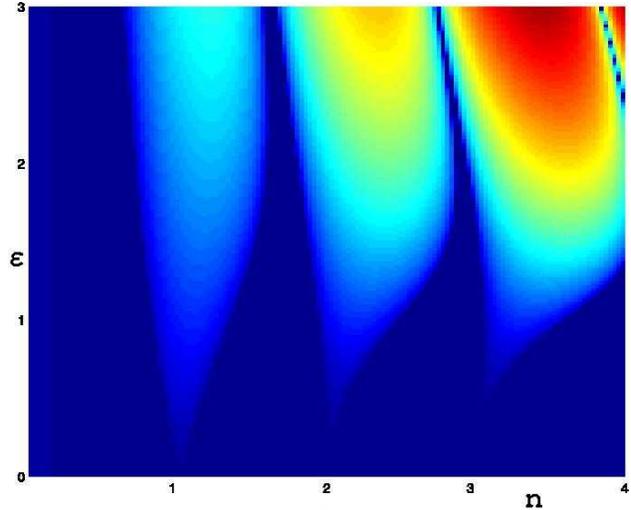}}
\caption{(Figures online colour at www.an-journal.org)
Stability regions for the Mathieu equation: on the horizontal axis there 
is the order of resonance, while the vertical axis describes the
strength
 of the coupling $\epsilon$. The blue region (the external one) is stable, the
 other colours represent the unstable solutions; the boundary between
 the two corresponds to the combinations of frequencies and $\epsilon$
 that give the steady-state solution (courtesy of V.\ Karas).}
\label{fig1}
\end{figure}
 An autoparametric system is one in which the time variation of the system
  parameters is not explicit, but depends on the coupling of the non-linear terms
   in different directions. In this context we will use the term parametric resonance
    to refer to both kind of systems. 
\subsection{Regions of non-linear resonance}
The first result which can be derived by using any  perturbative method is that
in the approximate solution  there are terms with the denominators
of the form $n \omega_r-m \omega_z$ (with $n$ and $m$ being integers).
These $n$ and $m$ cannot take any value, but they depend on the symmetry
 of the metric and of the perturbation: for a plane symmetric configuration one
  can demonstrate (e.g.\ Rebusco 2004) that $m=2\:p$ ($p$ integer).
  Due to this, the regions where $n \omega_r= 2 p \omega_z$ are candidate regions
   of parametric resonance (an analogous reasoning can be done for external resonance).
\ \  
These systems are well studied in engineering, where usually the main aim is to control the
 excited resonances: one can find examples of such couplings in aviation (e.g.\ the problem 
 of how to reduce the vibrations due to rotor blade flapping motion of helicopters), 
 building  (e.g.\ how the
  elevator ropes respond to high-rise buildings excitations), electronic circuits, etc.

 As it can be seen from Fig.~2 of Abramowicz (2005),
in General Relativity the epicyclic frequencies
are different ($\omega_z>\omega_r\:\: \forall r$) and there are specific radii at which they are commensurate: hence
at those radii an internal resonance between the radial and the vertical direction can
occur. The lower order resonance are stronger and more probable: the minimum
        value for $p$ is $p=1$, hence for a particle
        in a plane-symmetric strong field the resonance which is most likely to be
excited is the $3:2$ ($n=3$, because  $n=1,2$ correspond to $\omega_z \leq \omega_r$ ).
	 When the excitation of one mode reaches a critical amplitude, then the
	  linear response saturates, loses stability and the energy is transferred to 
	   the other mode and back.
	  Apart from the distance from the central object,
other conditions for the resonance are obviously that the damping does not avoid the
growth of the amplitudes and that the perturbation is not too strong (this would
lead the system to a chaotic state): these conditions are natural requirements.

In a real accretion disk the perturbation from the whole disk would get excited 
at the $r_{3:2}$ radius (radius at which the eigenfrequencies are in $3:2$ ratio) and
from there a wave would propagate through the whole disk (presumably with lower
amplitudes). 
A $3:2$ resonance was discovered in the planet Mercury: however in  that case
it is an external  resonance between the spin of the planet itself and its orbit
  around the Sun (Goldreich \& Peale $1966$).

\subsection{Frequency corrections and the Bursa line}
The solution which can be obtained with the method of Poincar\'e is a special solution, the so called
steady-state solution (constant amplitudes and frequencies):
this solution is such that 
 also if the frequencies are corrected, their ratio remains exactly $3:2$ (Hor\'ak 2004).
 On the plane ratio versus $\epsilon$ (see Fig.~1) the steady-state solution are the contours
  which divide the stable solutions (outer part) from the unstable ones (inner part).
   In a $j$-dimensional system instead of a line, there can exist a $j$-dimensional
    surface in which the solutions are steady-state.
  More intersetting solutions are those which are steady up to a certain order of approximation,
   but vary at higher order: these would be solutions close to the stable surface.
    We contemplated the possibility of a solution which is
    steady-state up to the third order, but which is allowed to vary starting from the
     next order. This would mean that the frequencies:
     \begin{eqnarray}
     \omega_r^*&=&\omega_r+\epsilon^2 \omega_r^{(2)}+\epsilon^3 \omega_r^{(3)}+O(\epsilon^4)\:\:\:,\nonumber\\ 
     \omega_z^*&=&\omega_z+\epsilon^2 \omega_z^{(2)}+\epsilon^3 \omega_z^{(3)}+O(\epsilon^4)
     \end{eqnarray}
     and the amplitude of oscillations are constant for $t\sim  \epsilon^{(-3)}$, but
      for longer times they change periodically (Hor\'ak 2004), exchanging
energy (amplitudes are anticorrelated, frequencies are related). 
 The dominant corrections $\omega_i^{(2)}$ depend quadratically on the amplitudes and
  they are due to the non-linearity. Hence for the observed frequencies
   we use the relation:
   \begin{equation}
   \omega_z^*=S \omega_r^*+Q ,
   \end{equation}
   where $S$ depends quadratically on the amplitudes of the perturbation. 
  In the case studied by Abramowicz et al.\ (2003) and Rebusco (2004), 
  $S=S(r_0, \delta r, \delta \theta, \alpha) $. In the 
   above mentioned papers the initial conditions of the perturbed geodesics were found
    in order to match the observed frequency-frequency slope of the neutron star 
    source Sco X-1. Indeed neutron star sources display different centroid frequencies
     in different observations, but amazingly they all fit on the same line, which is
      close to $3:2$. In the numerical study (Abramowicz et al.\ 2003) this was obtained
       by changing the initial radius $r_0$ and by using the $\alpha$ which gave the
        strongest response. In the analytical study (Rebusco 2004) the same result was
	 obtained by fixing $\alpha$ and varying the initial perturbations ($\delta r$ and
	  $\delta \theta$). The two procedures agree on the point that a weak perturbation
	    to a free particle leads to quasi-harmonic oscillations, whose
	     frequencies and ratio are close to the eigenfrequencies and to $3:2$ respectively
	      (see Fig.~2; we refer to the deviation from $3:2$ as the Bursa line).  
	      Consider the case of BHs and NSs: the mass of the first is greater, hence
	       $\omega_{\rm BH}<\omega_{\rm NS}$. Moreover,  
	    suppose that the entity of the perturbation in the two classes
  of objects is different: if it is stronger in NSs than in BHs, then both the
   frequency corrections and the observed ratio will be greater in NSs (they are
    proportional to the square of the amplitudes).\
     The weak point in is this model is that it cannot explain why the slope
      is $S=0.92$ for all NS sources,
      rather than a different value: with our simplifications we cannot discriminate
      the value of $S$, which depends on our choice concerning the perturbation. In principle
        we cannot even justify why the slope-slope relation is linear! 
       One may infer that this deals with the structure of NSs 
       themselves: in the equations that describe the disk oscillations, there must
        be a limit cycle, such that for different initial conditions the frequency
	 range changes, but at each time the frequencies fall on the same line.
	  Such limit cycle should be connected to the mass accretion rate in the inner
   part of the disk.

 We emphasize however that in the non-linear resonance model the frequencies
shift arises naturally, as well as the deviations from the exact commensurate ratio:
this  behaviour agrees with the observations. The actual challenge for theo\-re\-ticians is a full understanding of the ``Bursa line'', both qua\-litatively and quantitatively. 

\section{Qualitative discussion on other effects}
\subsection{Frequency-frequency slope: BHs and NSs}
 The strength of the perturbation is greater in NS than in BH sources: as a consequence
  the observed frequencies wander much more for NS than in BH
($\omega_{\rm NS}^{(2)}>\omega_{\rm BH}^{(2)}$ ).
\begin{figure}
\resizebox{\hsize}{!}
{\includegraphics[]{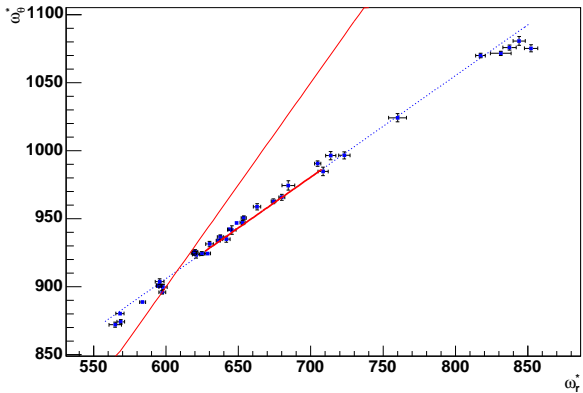}}
\caption{The dotted line is the least-squares best-fit to the data
points (the  observed kHz QPOs  frequencies in Sco X-1), which we call ``Bursa line''; the thin solid line corresponds to a slope
of $3:2$ (for reference) . The thick solid segment is the analytic approximation, in which
 the frequencies are scaled for comparison with observations.}
\label{fig2}
\end{figure}
However this explanation  is not completely satisfactory. The observed signal is not a direct
mirror of the disk oscillation, but one has to take into account the propagation and emission mechanisms.
In black holes the presence of the event horizon avoids any additional modulation: one can assume that the emitting flux has the characteristic frequencies
  of the most excited resonance ($r_{3:2}$), and that a subsequent modulation is just
   due to relativistic effects, such as strong gravitational lensing and
    Doppler effect (Bursa et al.\ 2004). Hence the timing properties of the observed 
    photons would be essentially distributed around  the frequency of oscillation of a single particle situated at a distance $r_{3:2}$ from the central object.
In neutron stars the modulation, which originates in the disk,  propagates
to the boundary layer (Abramowicz, Hor\'ak \& Klu\'zniak 2005;
Gilfanov, Revnivtsev \& Molkov 2003) and
there the original frequencies of oscillation (the ones at $r=r_{3:2}$) are subject to
the influence of the Roche lobe overflow: eventually in this case it is
difficult to predict
how the observed frequencies are related to the original ones (see Hor\'ak 2005).
       
\subsection{The effect of a periodic forcing}
It is widely accepted that in  NS sources the QPO frequencies  depend on the spin: this
 indicates that the the NS spin directly excites the disk oscillations
  (see Lee 2005 and Sramkova 2005 
for numerical results). Let us add external forcing terms to 
   our equations: $A \cos{\omega_0 t}$ and $B \cos{\omega_0 t}$. In order to
    study them with a perturbative method it is convenient to
	     introduce $\tilde A$ and $\tilde B$ such that $A=\epsilon^j \tilde A$ and $B=\epsilon^k \tilde B$ ($j$ and $k$ are integers: they can be different to
	      take into account a different forcing in the two directions).

By considering this forcing a new whole range of resonance conditions is possible:
     \begin{equation}
	      n\omega_z \pm m \omega_r=q \omega_0,
     \end{equation}
with $n,m,q=0,1,2,...\,$. An interesting example was studied by Klu\'zniak et al.\ (2004): indeed the combination of frequencies such that the QPOs are
separated by $\frac{1}{2}$  the spin frequency would explain the 
recent observations of the millisecond pulsar
SAX J1808.4-3658 (Wijnands et al.\ 2003).
Another possibility in agreement with the data from NS, is that
the difference of the QPOs peaks remains locked at the spin frequency itself.
 
 \subsection{MRI Turbulence}
In the analytical study the turbulence terms were set to zero:
up to now only a numerical
study of these stochastic differential equations (SDEs) have been done.
 Anyway from the first experiments (private communication with R.\ Vio and H.\ Madsen) we can
 already see that if a Gaussian noise is assumed, its standard deviation cannot
 be too large, or the resonance regime will be disrupted in favour of a chaotic
 one. However for small standard deviations the turbulence introduces new resonances
 and it drives the system from one stable resonance to the other.
 The study of these SDEs is very promising:  it is necessary to take into
account  these terms in order to compare our model with observations. 

The stochastic terms permit to analyze new behaviours and hopefully their study will lead to a better understanding of the MRI itself.
     
\section{Conclusions}
In weakly non-linear oscillators the leading terms are periodic, with  frequencies close 
 to the eigenfrequencies of the system: the frequency shift is proportional to the square
of the amplitude of the perturbation. The internal coupling of two or more subsystems 
 introduces the possibility
 of parametric resonances to occur (in the regions where the characteristic frequencies are commensurate). Another feature of non-linear resonance is the presence
   of  subharmonics. 

All this suggests that the observed kHz QPOs are a non-linear phenomenon. With the simple toy-model
            of a perturbed test particle in the strong gravitational field of compact objects, we
	     can qualitatively explain the position of the twin QPO peaks in a frequency-frequency plane.
	      However many puzzles remain to be solved: is there a limit cycle in the system?
	       Is the presence of the boundary layer sufficient to justify the differences between NSs and BHs?
	        Are the shift and the slope of the Bursa line related? How does turbulence affect the resonance
		 conditions? Attacking these questions with the different instruments of theory, computation and observations may lead to a better understanding of the physics of accretion, of
  the mathematics of non-linear systems, and to the test of strong gravity itself.

\acknowledgements Pragne podziekowac Markowi  Abramowicz za giving me
 the possibility to work on such a fascinating subject and to meet regularly the research group.
 I would thank him, Wlodek Klu\'zniak, Jiri Hor\'ak and Roberto Vio for the helpful discussions and suggestions and Vladimir Karas for the hospitality at the Astronomical Institute in Praha and for making the plot in Fig.~1. I am grateful to Axel Brandenburg for helping me in the editing.

\end{document}